\documentclass{article}
\usepackage{spconf,amsmath,graphicx}
\usepackage{booktabs}

\title{Direct Noisy Speech Modeling for Noisy-to-Noisy Voice Conversion}
\name{Chao Xie, Yi-Chiao Wu, Patrick Lumban Tobing, Wen-Chin Huang and Tomoki Toda}
\address{Nagoya University, Japan}
\begin{document}
\ninept
\maketitle
\begin{abstract}
Beyond the conventional voice conversion (VC) where the speaker information is converted without altering the linguistic content, the background sounds are informative and need to be retained in some real-world scenarios, such as VC in movie/video and VC in music where the voice is entangled with background sounds. As a new VC framework, we have developed a noisy-to-noisy (N2N) VC framework to convert the speaker's identity while preserving the background sounds. Although our framework consisting of a denoising module and a VC module well handles the background sounds, the VC module is sensitive to the distortion caused by the denoising module. To address this distortion issue, in this paper we propose the improved VC module to directly model the noisy speech waveform while controlling the background sounds. The experimental results have demonstrated that our improved framework significantly outperforms the previous one and achieves an acceptable score in terms of naturalness, while reaching comparable similarity performance to the upper bound of our framework.  
\end{abstract}
\section{Introduction}
\label{sec:intro}

Voice conversion (VC) is a technique for converting the voice characteristics of a source speaker into that of a target speaker without changing the linguistic contents. With the advent of deep learning in recent years, neural network-based methods have taken center stage in VC studies, bringing about tremendous improvements. The state-of-the-art method \cite{liu2020non}, according to the latest Voice Conversion Challenge (VCC) 2020 \cite{zhao2020voice}, shows comparable similarity to the natural target speech with only a slight disparity for naturalness. 

In addition to the conventional speaker VC, the VC techniques have been further applied to more challenging applications such as dubbing \cite{mukhneri2020voice},  singing \cite{9054199}, \cite{sisman2019singan}, \cite{doi2012singing} and data augmentation \cite{shahnawazuddin2020voice}, whose background sounds should be preserved. Take VC for movies as an example, converting the actor's voice characteristics entangled with speech content and background sounds is hard. Although suppressing the background sounds is helpful to reduce the interference, the background sounds should be well retained. The same situation is also encountered in VC-based data augmentation for automatic speech recognition (ASR), where suppressing the background noise can improve the VC performance. Nevertheless, the background noise is also a valuable resource to be retained for improving the robustness of the downstream system. Therefore, flexibly dealing with the background sounds in the noisy VC tasks is essential.

\begin{figure}[t]
\begin{minipage}[b]{1.0\linewidth}
  \centering
  \centerline{\includegraphics[width=\linewidth]{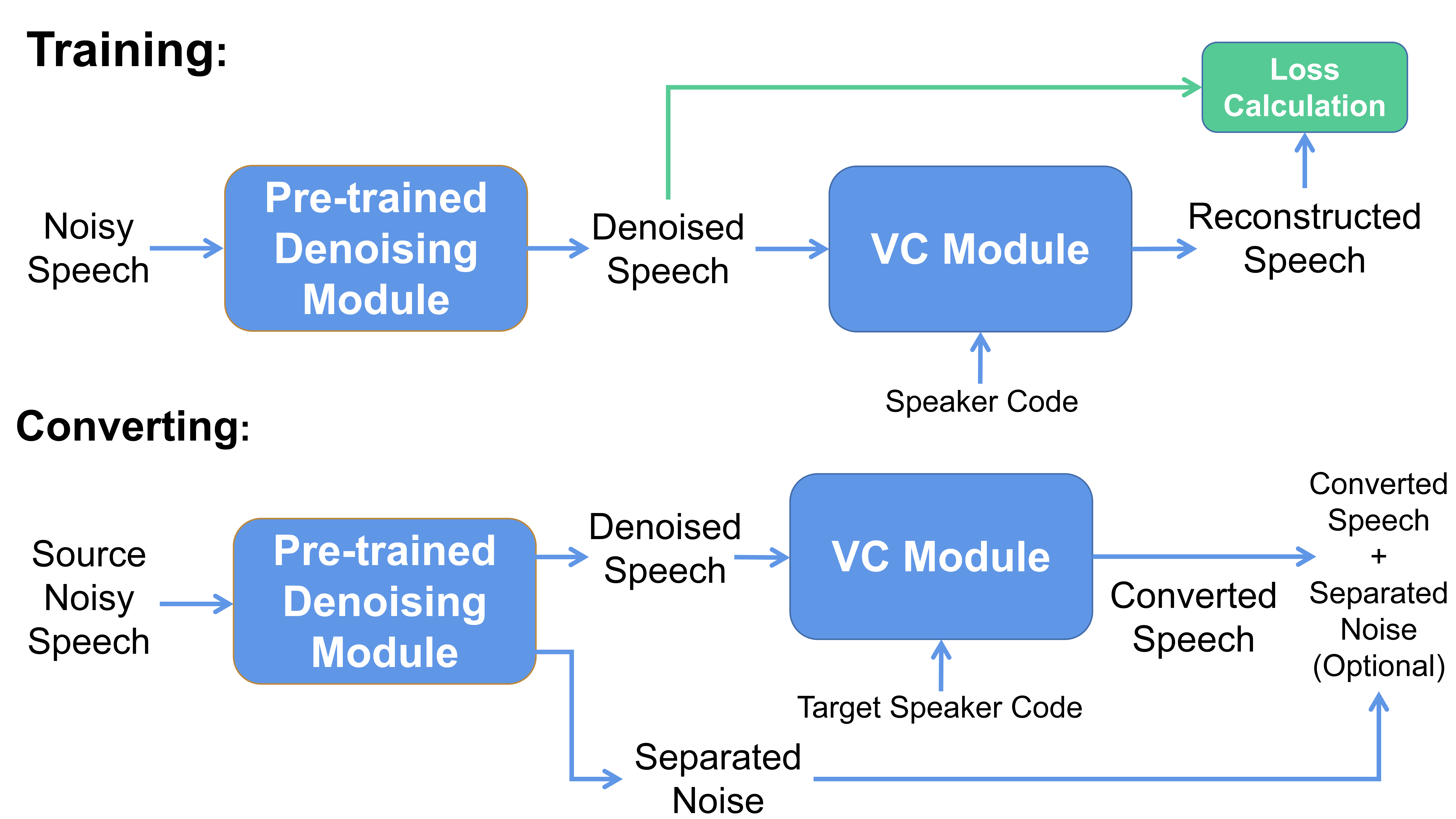}}
  \centerline{(a) Baseline framework}\medskip
\end{minipage}

\begin{minipage}[b]{1.0\linewidth}
  \centering
  \centerline{\includegraphics[width=\linewidth]{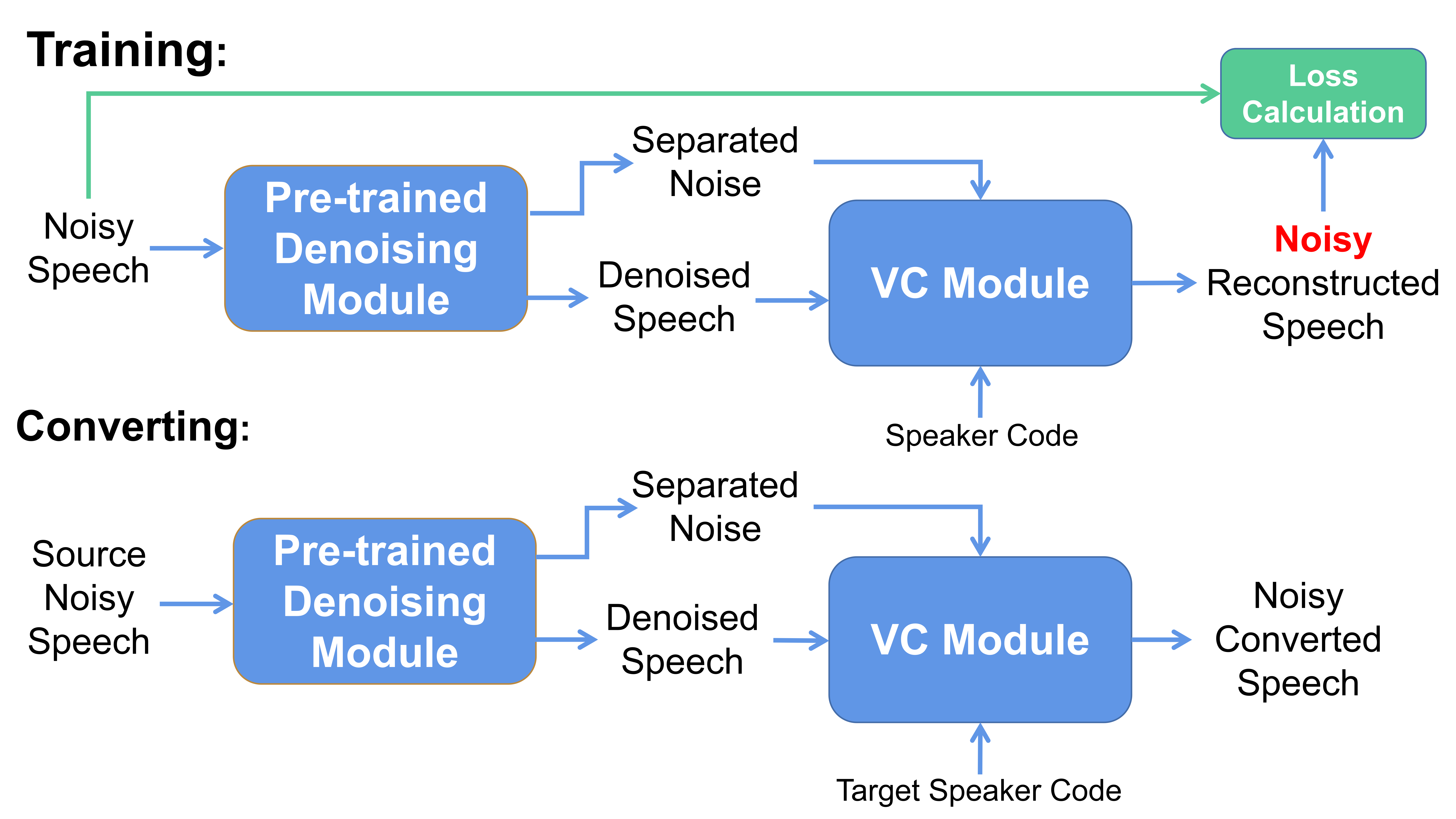}}
  \centerline{(b) Proposed framework}\medskip
\end{minipage}

\caption{Overview of N2N VC frameworks.}
\label{fig:overflow}
\end{figure}

The majority of previous works focus on noise-robust VC, where the background sounds are discarded as interference. Takashima {\it et al.} \cite{takashima2013noise} proposed a sparse representation-based VC using non-negative matrix factorization to filter out the noise. Miao {\it et al.} \cite{miao2020noise} presented a noise-robust VC that introduced two filtering methods into the preprocessing stage and the post-processing stage, respectively, to suppress the noise. Mottini {\it et al.} \cite{mottini21_ssw} proposed a denoising autoencoder-based zero-shot VC. One of the biggest limitations of these previous methods was that both noisy and clean source/target speech data were basically required. 

As another work, Hsu {\it et al.} \cite{hsu2019disentangling} proposed a text-to-speech (TTS) based VC method to handle the noisy speech data while controlling the noise. They augmented the clean training set with a copy that mixed with the noise clips but reused the same transcript and speaker label, based on which a variational autoencoder was trained to learn the disentangled representations of the speaker identity and background noise. Adversarial training was conducted to improve the degree of disentanglement further. These two techniques were used in the training of the TTS model to control the speaker identity and the background noise. However, clean VC data was still necessary, and the quality of the background noise in the converted speech was quite limited.  

To meet new requirements for the noisy VC tasks, we have developed a noisy-to-noisy (N2N) VC framework, where the speaker conversion is achieved while retaining the background noise \cite{xie2021noisy}. The first "noisy" signifies only noisy source/target speech data are available for VC training; the second "noisy" means the background sounds are preserved and controllable. Inspired by the previous work \cite{ValentiniBotinhao2016InvestigatingRS} where a TTS model was combined with a recursive neural network-based denoising method to deal with the noisy data, we have introduced a state-of-the-art denoising module into the N2N VC framework. Although the developed framework is capable for the noisy VC tasks, we have found that the denoising module tends to bring in the distortion problem significantly degrading the VC performance.

In this paper, to tackle this distortion problem, we propose the modified VC module to directly model the noisy speech that is free from the distortion caused by the denoising module. The effectiveness of the proposed method is investigated by conducting objective and subjective evaluations. The experimental results show that the proposed method significantly outperforms the previous one and achieves good conversion performance with well-retained background sounds\footnote{Samples can be found at https://github.com/chaoxiefs/n2nvc.}.

\section{Baseline N2N VC Framework}
\label{sec:baseline}

Figure~\ref{fig:overflow} (a) illustrates an overview of the baseline framework. This framework consists of a denoising module and a VC module. The denoising module is developed beforehand with publicly available datasets. It is utilized for separating a noisy speech waveform into speech and noise waveforms in the time-domain:
\begin{equation}
\label{eq:separate}
\mathbf{n} = \mathbf{y} - \mathbf{d},
\end{equation}
where $\mathbf{y} = \left\{y_{1}, \ldots, y_{T}\right\}$ denotes the noisy speech waveform, $\mathbf{d} = \left\{d_{1}, \ldots, d_{T}\right\}$ denotes the denoised speech waveform estimated by the denoising model, and $\mathbf{n} = \left\{n_{1}, \ldots, n_{T}\right\}$ denotes the separated noise waveform.

The VC module is trained with the denoised speech. In the conversion stage, the denoising module separates input noisy speech into speech and noise, and only the denoised speech is delivered to the VC module to generate converted speech. The separated noise can be either discarded or superimposed to the converted speech.

\subsection{Denoising Module}
\label{ssec:denoising module}

\begin{figure}[t]
\begin{minipage}[b]{1.0\linewidth}
  \centering
  \centerline{\includegraphics[width=\linewidth]{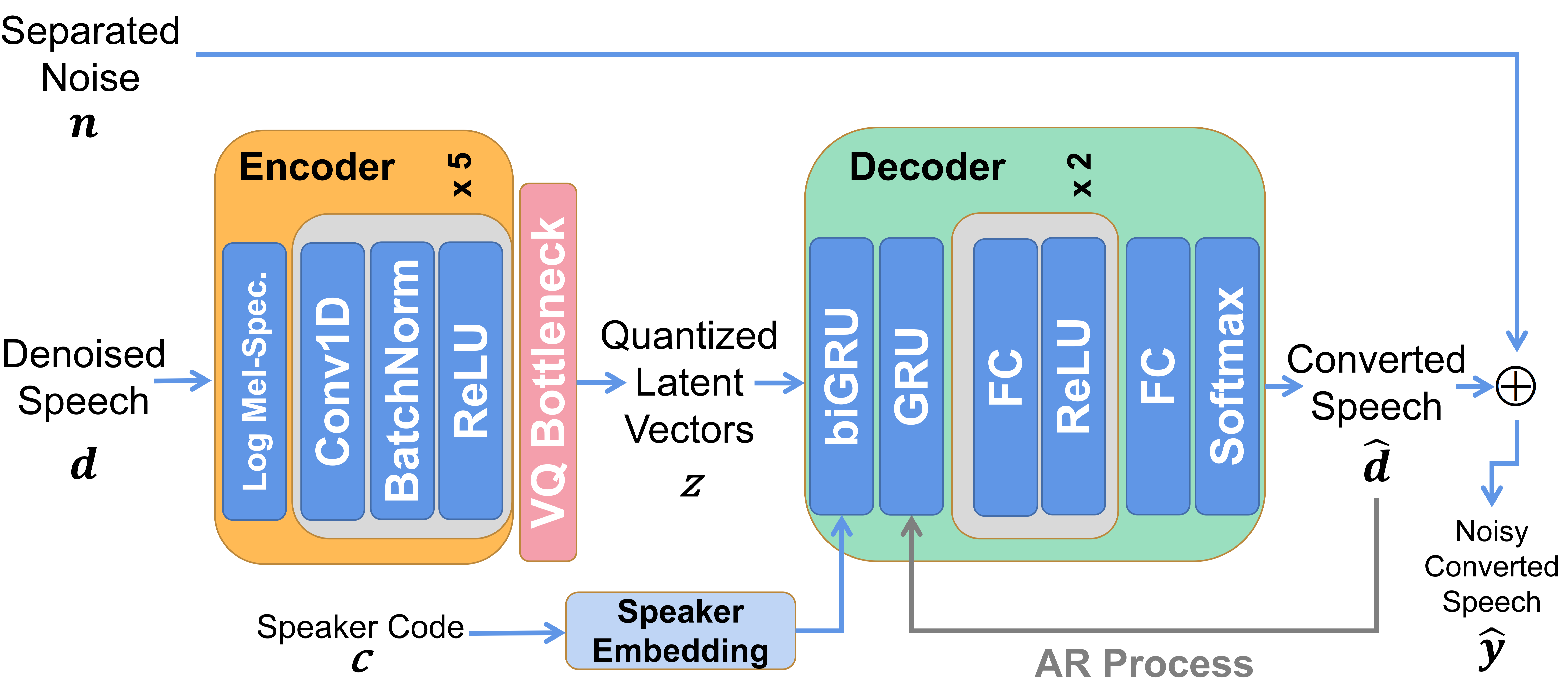}}
  \centerline{(a) VQ-VAE in the baseline framework}\medskip
\end{minipage}

\begin{minipage}[b]{1.0\linewidth}
  \centering
  \centerline{\includegraphics[width=\linewidth]{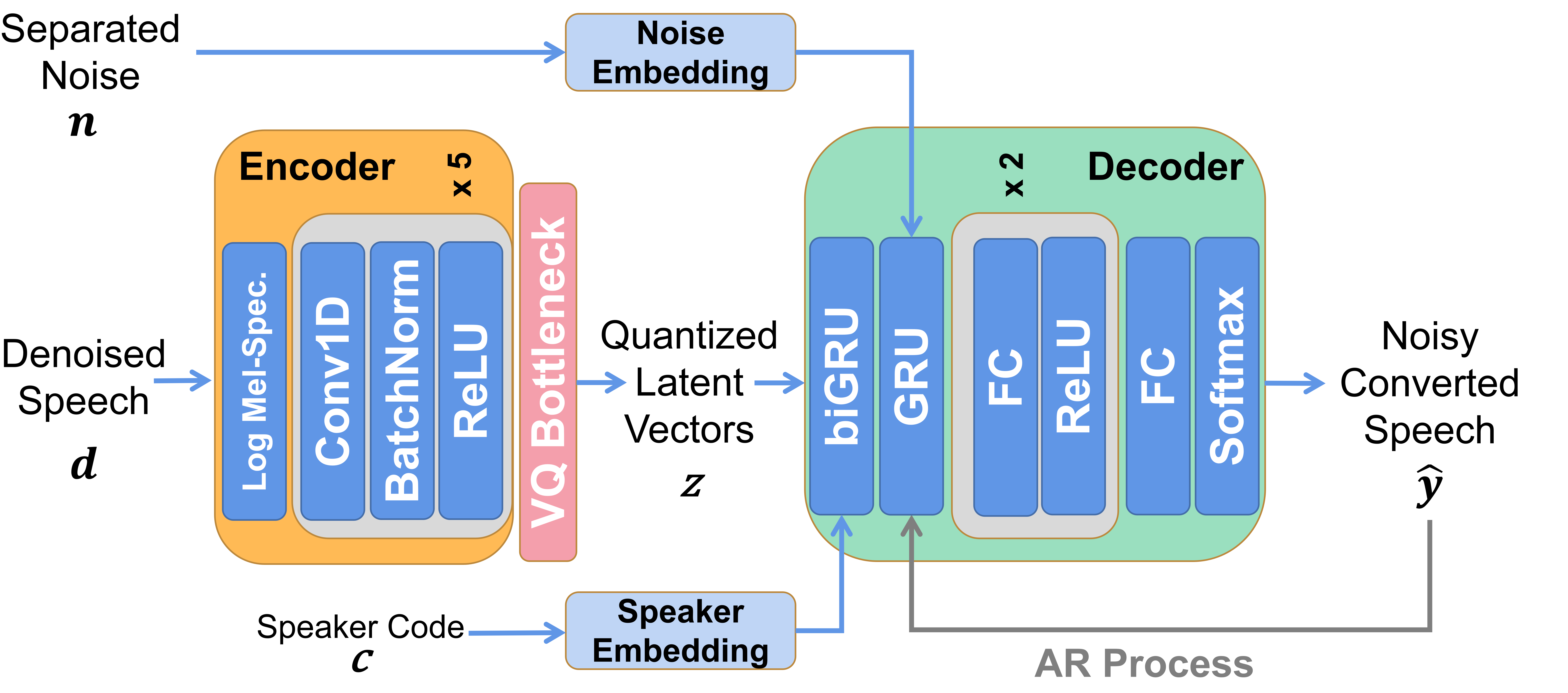}}
  \centerline{(b) Modified VQ-VAE in the proposed framework}\medskip
\end{minipage}

\caption{Conversion flow with VQ-VAEs in the baseline and proposed frameworks.}
\label{fig:vqvae}
\end{figure}

The denoising module is implemented as Deep Complex Convolution Recurrent Network (DCCRN) \cite{hu2020dccrn}, which ranked first for the real-time-track in Deep Noise Suppression (DNS) Challenge 2020 \cite{reddy2021icassp}. DCCRN is a convolutional recurrent network (CRN) based single-channel denoising method capable of handling the complex values. More details about DCCRN can be found in \cite{hu2020dccrn}.

In our work, since DCCRN is utilized as a separation model conducting the separation operation as shown in Eq.~(\ref{eq:separate}), the power of the denoised speech should be matched to the clean reference speech. Hence, the original scale-invariant signal-to-noise ratio (SI-SNR) loss used in DNS Challenge is substituted by the scale-dependent signal-to-distortion (SD-SDR) loss \cite{le2019sdr}.

\subsection{VC Module}

A vector-quantized variational autoencoder (VQ-VAE) based non-parallel VC method \cite{van2020vector} is implemented as the VC module, the structure of which follows \cite{xie2021noisy}. Figure~\ref{fig:vqvae} (a) illustrates the structure of VQ-VAE in the baseline. The log mel-spectrogram extracted and processed by the encoder is sent to a vector-quantized bottleneck to discard speaker information. The decoder adopts a lightweight recurrent network based on WaveRNN \cite{2018wavernn} to reconstruct the waveform in an autoregressive (AR) manner that generates the current sample based on the previously generated ones. The VQ-VAE is trained with the denoised speech waveforms of multiple speakers. The VQ-VAE loss function consists of an encoder and bottleneck term and a decoder term, where the decoder term is based on the joint probability distribution $p\left(\mathbf{d} \mid \mathbf{c}, \mathbf{z} \right)$ factorized into the conditional probability distributions at each time step as follows:
\begin{equation}
\label{eq:baseline}
p\left(\mathbf{d} \mid \mathbf{c}, \mathbf{z} \right)=\prod_{t=1}^T p\left(d_{t} \mid d_{1}, \ldots, d_{t-1}, \mathbf{c}, \mathbf{z}\right),
\end{equation}
where $\mathbf{z}$ and $\mathbf{c}$ are defined as a quantized latent vectors from the bottleneck and a speaker code corresponding to the denoised speech $\mathbf{d}$, respectively. \cite{oord2017neural} presents detailed explanation about VQ-VAE.

\subsection{Drawbacks of the Baseline}
Although our previous experimental results \cite{xie2021noisy} have shown the effectiveness of the baseline on N2N VC tasks, there is still a significant gap from the upper bound of the framework, that is, VQ-VAE directly trained with the clean dataset. The main reason is that the VQ-VAE in the baseline is trained to model the denoised speech waveform suffering from distortion caused by the denoising model. This distortion is further propagated to the VC module, causing quality degradation of the converted speech. Besides, the baseline can not achieve noisy converted speech generation in one stage: the clean converted speech needs to be generated first, and then, the separated noise is superimposed in the time domain.

\section{Proposed N2N VC Framework}
\label{sec:improved}

Within all the data involved in the baseline method, only the noisy speech is free from the distortion caused by the denoising process. Therefore, we propose a method to directly model the noisy speech to alleviate the distortion issue and also make it possible to achieve one-stage noisy converted speech generation.

To derive the joint probability distribution of the noisy speech, we transform the joint probability distribution of the denoised speech given in Eq.~(\ref{eq:baseline}) by changing a random variable from $\mathbf{d}$ to $\mathbf{y}$ with their linear relationship given in Eq.~(\ref{eq:separate}) as follows:
\begin{equation}
p\left(\mathbf{d} \mid \mathbf{c}, \mathbf{z} \right) = p\left(\mathbf{y} \mid \mathbf{n}, \mathbf{c}, \mathbf{z} \right) \qquad \mbox{s.t.} \quad \mathbf{y} = \mathbf{d} + \mathbf{n},
\end{equation}
where the separated noise $\mathbf{n}$ is regarded as a given variable. Because this formulation still suffers from the distortion due to the explicit constraint $\mathbf{y} = \mathbf{d} + \mathbf{n}$, we remove it. Then, as in Eq.~(\ref{eq:baseline}), the unconstrained joint probability distribution is factorized into the conditional probability distributions at each time step as follows:
\begin{equation}
p\left(\mathbf{y} \mid \mathbf{n},  \mathbf{c}, \mathbf{z} \right)=\prod_{t=1}^T p\left(y_{t} \mid y_{1}, \ldots, y_{t-1}, n_{1}, \ldots, n_{t}, \mathbf{c}, \mathbf{z}\right),
\end{equation}
where we further assume that $y_t$ doesn't depend on succeeding samples of the separated noise since this assumption makes it possible to achieve a low-latency real-time conversion process. An overview of the proposed framework is shown in Fig.~\ref{fig:overflow} (b). Both the denoised speech and the separated noise are used to train the VC module. Note that the denoised speech is still used as an input of the VQ-VAE encoder. As shown in Fig.~\ref{fig:vqvae} (b), VQ-VAE is modified by introducing the separated noise as a condition embedded in the decoder. We expect that the modified VQ-VAE will implicitly learn the relationship among noisy speech, separated noise, and denoised speech.

In the conversion stage, the modified VQ-VAE directly generates the noisy converted speech in the AR process with the target speaker code, the quantized latent vectors extracted from the source denoised speech, and the separated noise from the source noisy speech. Note that this model is also capable of generating the clean converted speech by setting the separated noise to a zero sequence.

\begin{figure}[t]
  \centering
  \includegraphics[width=\linewidth]{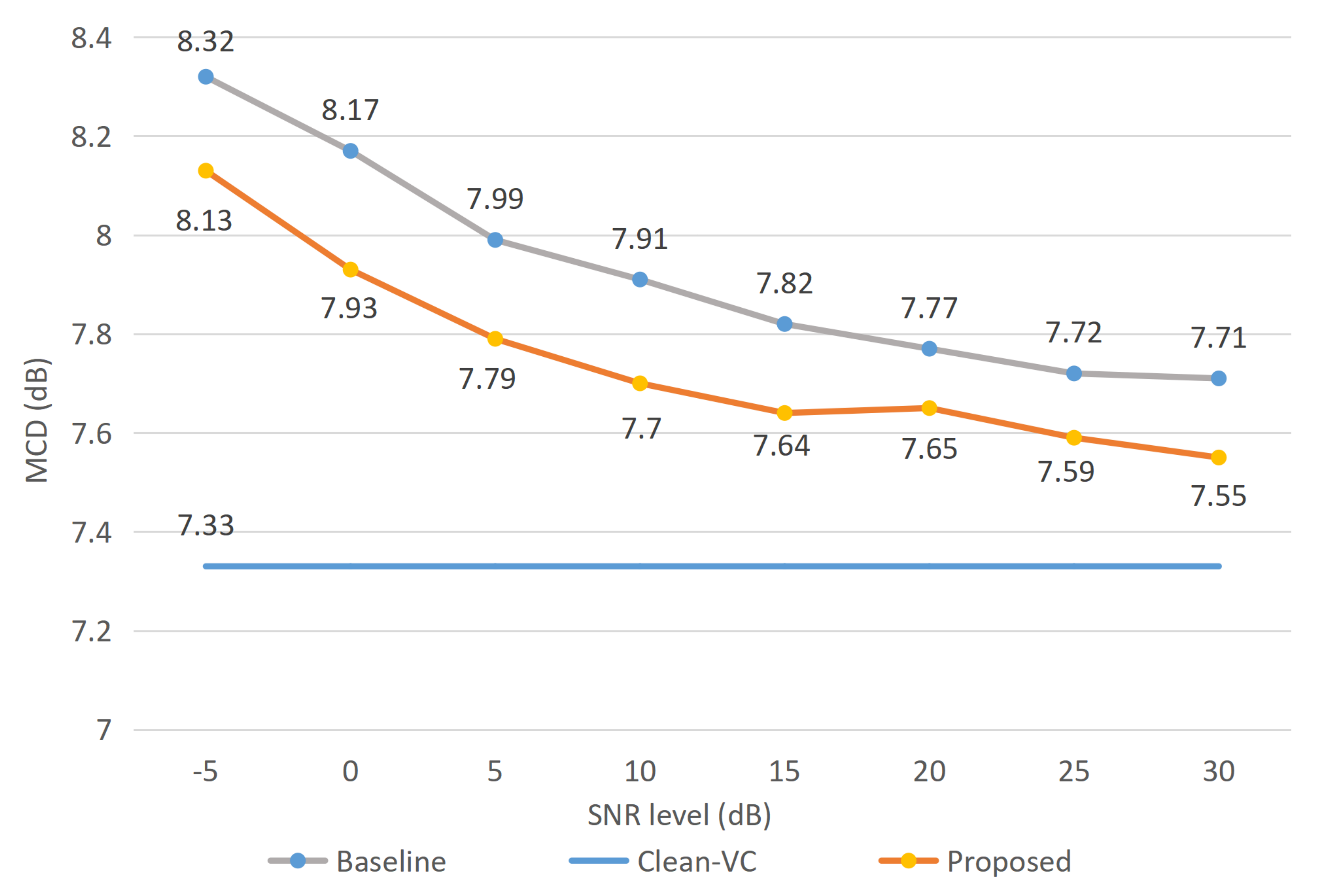}
  \caption{MCD as a function of SNR level.}
  \label{fig:mcd}
\end{figure}

\section{EXPERIMENTAL EVALUATIONS}
\label{sec:exp}

\subsection{Experimental Conditions}
\label{subsec:exp_condition}

The experimental evaluations were conducted on 8 kHz sampled speech data since we focused on telephone speech conversion and data augmentation for ASR of telephone speech.

DNS Challenge 2020 dataset \cite{reddy2021icassp} was used to train DCCRN. It contained 500 hours of speeches from 2,150 speakers in various languages and a total of 65,000 noise clips in 150 classes. 6,000 speech clips and 500 noise clips were randomly sampled as the validation data. The SNR levels (dB) were set from 5 to 20.

VCC 2018 dataset \cite{lorenzo2018voice} and PNL 100 Nonspeech Sounds \cite{hu2010tandem} were mixed to construct the noisy VC dataset. Both of them were unseen for the denoising module. The noisy training set was built by mixing the speech clips from all 12 speakers (972 utterances in total) in the VCC training set with the noise clips sampled from N1 to N85 (85 clips in 9 categories) in the PNL 100 at certain SNR levels (dB): (6, 8, 10, 12, 14, 16, 18, 20); while the noisy evaluation set was composed of the speech clips from 4 source speakers (VCC2SM3, VCC2SM4, VCC2SF3, VCC2SF4) and 2 target speakers (VCC2TF2, VCC2TM2) in the VCC evaluation set mixed with the noise clips from N86 to N100 (15 clips in 11 categories) in the PNL 100. It needed to be emphasized that the noise categories in the evaluation set were not included in the training set.

For the objective evaluation, 8 noisy evaluation sets were built parallelly at SNR levels (dB) (-5, 0, 5, 10, 15, 20, 25, 30). In each dataset, a speaker had 35 noisy utterances. The same utterance from the different speakers among different datasets had the same noise category. As for the subjective evaluation, two evaluation sets at SNR 7 dB and 15 dB were leveraged. They were parallelly constructed in the same way as the objective evaluation. In each daatset, 5 converted speeches with different noise categories were randomly sampled per conversion pair. In general, each method provided 80 samples to be tested in the subjective evaluation. The corresponding noisy target speeches were leveraged as the ground truth, which involved 20 in total.

For the training of DCCRN and VQ-VAE, we used the same settings as we had done in \cite{xie2021noisy}.

\begin{figure}[t]
  \centering
  \includegraphics[width=\linewidth]{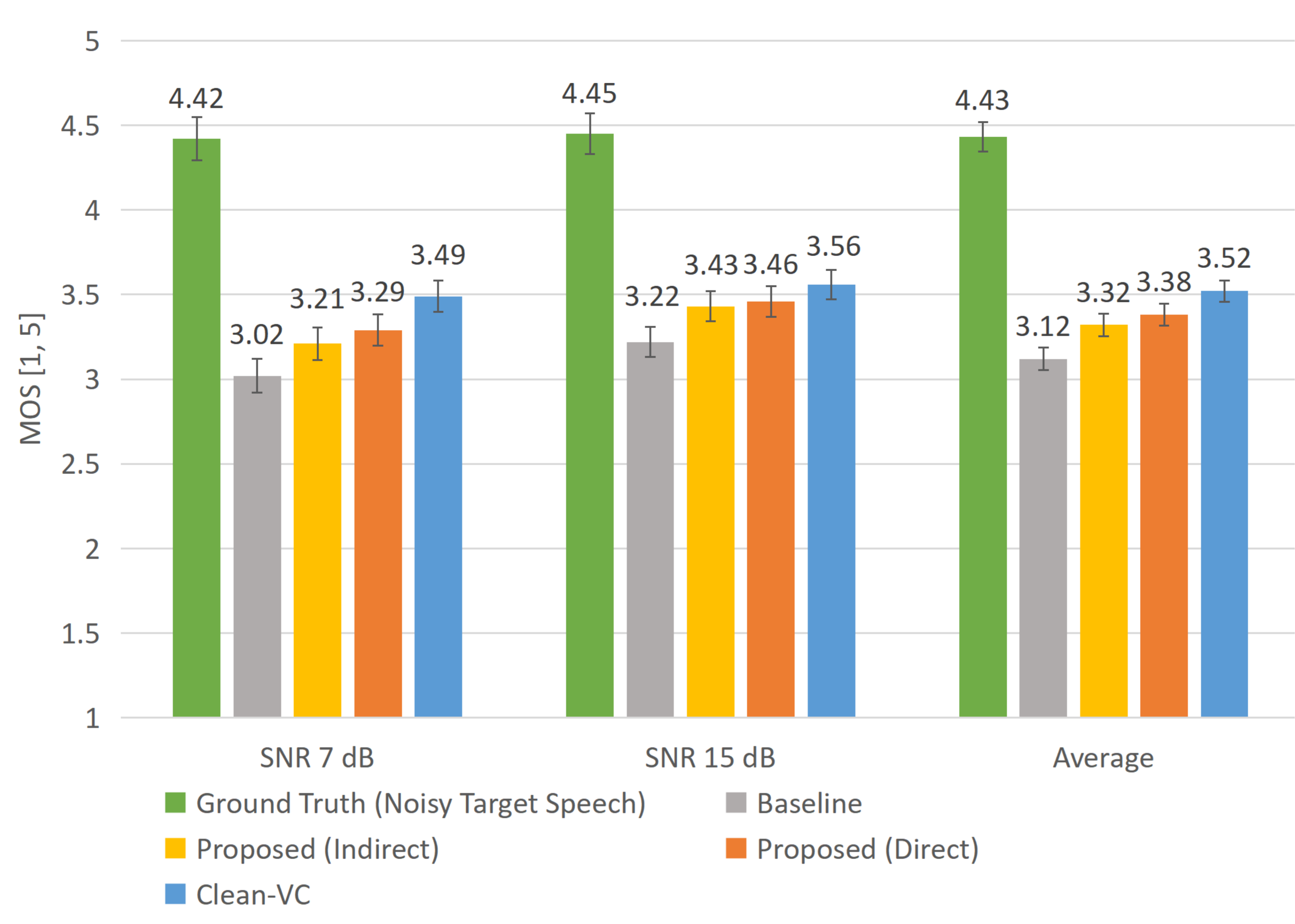}
  \caption{Result of the subjective evaluation on naturalness. Error bars show 95\% confidence intervals.}
  \label{fig:mos}
\end{figure}

.
\subsection{Experimental results}
Our proposed method was compared with the baseline detailed in Section~\ref{sec:baseline} and the upper bound of our framework: VQ-VAE trained on the clean VCC dataset denoted as Clean-VC. The evaluation of DCCRN on the denoising task was conducted in \cite{xie2021noisy}.

\subsubsection{Objective evaluation}
Mel cepstral distortion (MCD) \cite{kubichek1993mel} was employed as the objective measurement. Clean evaluation reference was leveraged. Therefore, the baseline did not superimpose the separated background noise. For our method, the noise condition was set to zero sequences to generate clean converted speech.

Figure~\ref{fig:mcd} demonstrates the results of MCD on the evaluation sets mentioned in Section~\ref{subsec:exp_condition}. As the upper bound, Clean-VC achieves the best score of 7.33.  The proposed method significantly outperforms the baseline under all SNR levels. When the SNR level goes up, making residual interference less in the denoised data, the proposed method achieves a better MCD score. The same trends are observed in the baseline.

\begin{figure}[t]
\begin{minipage}[b]{1.0\linewidth}
  \centering
  \centerline{\includegraphics[width=\linewidth]{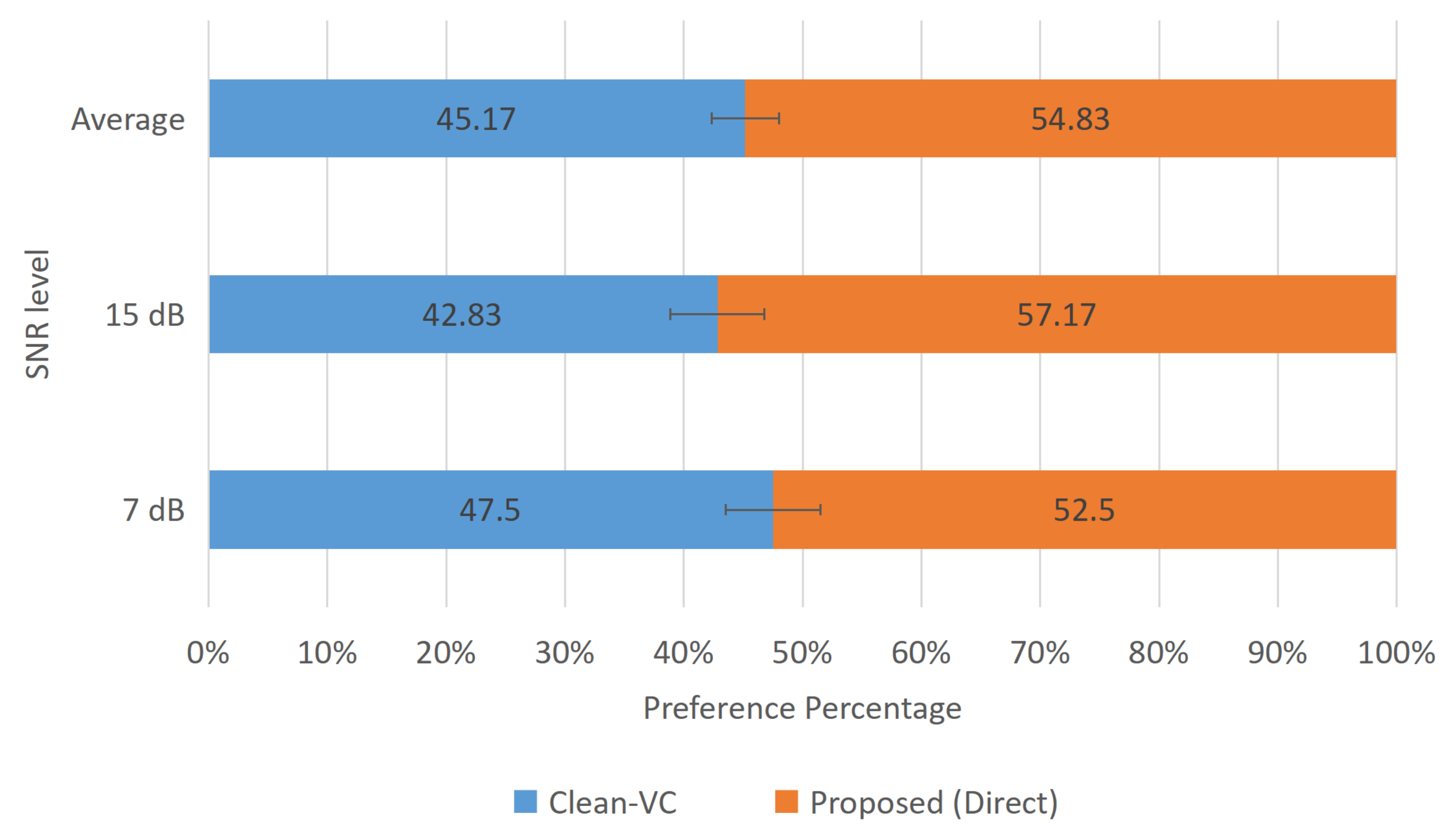}}
  \centerline{(a) Preference scores between the Clean-VC and the Proposed (Direct)}\medskip
\end{minipage}

\begin{minipage}[b]{1.0\linewidth}
  \centering
  \centerline{\includegraphics[width=\linewidth]{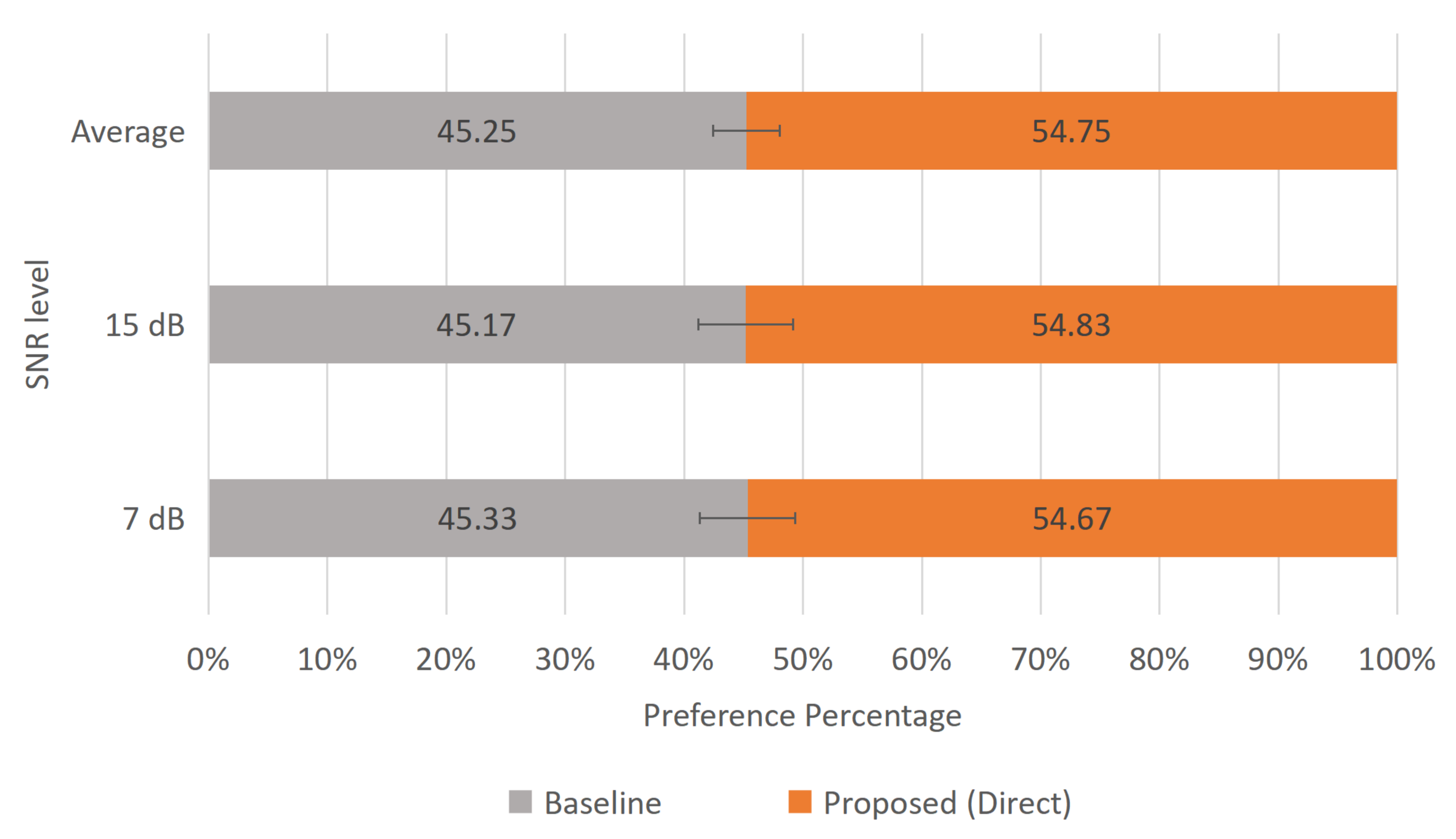}}
  \centerline{(b) Preference scores between the baseline and the Proposed (Direct)}\medskip
\end{minipage}

\caption{Preference scores with 95\% confidence intervals as the results of XAB tests on speaker similarity.}
\label{fig:xab}
\end{figure}

\subsubsection{Subjective evaluation}
The subjective evaluation was conducted as the mean opinion score (MOS) test for measuring the naturalness of the converted speech and the XAB tests for comparing the performance on similarity among different methods. Fifteen listeners participated in the subjective evaluation.

As we aimed at noisy-to-noisy VC, the noisy converted speeches were involved in the subjective evaluation. Specifically, the samples from the baseline were superimposed with the separated noise, denoted as {\bf Baseline}, while the converted samples of the Clean-VC were superimposed with the corresponding original noise clips, denoted as {\bf Clean-VC}. The proposed method, denoted as {\bf Proposed (Direct)}, synthesized the noisy converted speech directly. To further evaluate the proposed method, we set another group denoted as {\bf Proposed (Indirect)}, where the clean converted speeches were generated first and then superimposed with the separated noise.  

In the MOS test, the participants were required to give an overall naturalness score for both the speaking voice and the background sounds from 1 to 5 (higher is better). The category of the background sounds and its original record clip were given as the reference to assist them in judging its quality. 

The results of the MOS test are shown in Fig.~\ref{fig:mos}. The ground truth gets a MOS value of 4.43, indicating that the participants could give reliable evaluations for the noisy speech samples. The baseline and the proposed methods achieve better MOS values at a higher SNR level. The upper bound, {\bf Clean-VC}, gets 3.52 on average. {\bf Proposed (Direct)} achieves 3.38 on average, significantly outperforming {\bf Baseline} with an average score of 3.12. At SNR 15 dB, {\bf Proposed (Direct)} achieves the MOS value of 3.46 with a small gap from {\bf Clean-VC} with 3.56, while a large margin is still observed for {\bf Baseline} with 3.22. These results have proved the effectiveness of the proposed method. Moreover, {\bf Proposed (Direct)} and {\bf Proposed (Indirect)} achieves similar average scores of 3.38 and 3.32, respectively, which has shown the proposed one-stage noisy conversion does not degrade the quality of the background sounds. 

In the XAB tests, the participants listened to a target noisy speech as the reference and two noisy converted speeches from two methods to select which one sounded more likely from the same speaker of the reference. Since {\bf  Proposed (Direct)} got slightly higher MOS values than {\bf Proposed (Indirect)}, only the former was evaluated with {\bf Clean-VC} and {\bf Baseline}. 

The results of the XAB tests are shown in Fig.~\ref{fig:xab}. {\bf Clean-VC}, {\bf Baseline}, and {\bf Proposed (Direct)} receives similar preference percentage around 50\% under both SNR 7 dB and 15 dB. These results have shown that the proposed framework is competitive with the upper bound for the performance on similarity.

\section{Conclusion}

In this paper, we have presented a noisy-to-noisy VC framework that can only rely on noisy VC training data and is capable of controlling the background noise in the converted speech. This framework is composed of a denoising module and a VC module. The denoising module is utilized to separate noisy input speech into speech and noise before sending the data to the VC module. To further improve the performance, we have successfully modified the VC module by embedding the separated noise as the condition to model the noisy speech directly. Both objective and subjective evaluations have proved that the proposed method effectively bridges the margin on the naturalness between the baseline and the upper bound, and is competitive with the upper bound for the similarity. In the future, we will further investigate the controllability of the noise condition and improve the VC performance. 

{\bf Acknowledgement:} This work was partly supported by JST CREST Grant Number JPMJCR19A3, Japan.

\vfill\pagebreak


\bibliographystyle{IEEEbib}
\bibliography{strings,refs}

\end{document}